\documentclass{iaus}
\usepackage{graphicx}

\title[First results of the ACS LCID project] 
{The ACS LCID project:\\ overview and first results}

\author[Gallart]   
{Carme Gallart$^1$, for the LCID Team
  \thanks{LCID: Local Cosmology from Isolated Dwarfs; http://www.iac.es/project/LCID }}

\affiliation{$^1$Instituto de Astrof\'\i sica de Canarias, C/ V\'\i a L\'actea s/n, 38200 La Laguna, Tenerife, Spain. \break email: carme@iac.es\\[\affilskip]}

\pubyear{2007}
\volume{241}  
\pagerange{1}
\date{}
\jname{Stellar Populations as Building Blocks of Galaxies}
\editors{A. Vazdekis \& R. Peletier, eds.}
\begin{document}

\maketitle

\begin{abstract}
The star formation history (SFH) of Local Group galaxies is a powerful
tool for studying their evolution, including chemical enrichment
histories and stellar population gradients, which in turn may shed
light on the role of reionization or Supernovae feedback in galaxy
formation and evolution. In particular, isolated dwarfs are ideal
laboratories since their evolution has not been complicated by the
vicinity of giant galaxies. In this paper, we present the project {\it
Local Cosmology from Isolated Dwarfs (LCID)}, aimed at deriving
detailed SFHs for a sample of Local Group isolated dwarf galaxies. To
accomplish this goal we have collected, using the ACS on board the
HST, color-magnitude diagrams (CMD) reaching their oldest main
sequence turnoff ($V\simeq28$ or M$_V \simeq +$3.5) with good
photometric accuracy.  Some preliminary results from the CMDs are
shown and briefly discussed.
\end{abstract}

\firstsection 
\section{Motivation and aims of the project}

\noindent

The SFH of Local Group (LG) galaxies can be studied very accurately
from CMDs reaching the oldest main sequence turnoffs (ideally
complemented from spectroscopy from individual stars). It provides key
details on the processes of formation and evolution of these galaxies
and is, therefore, a complementary path to high-redshift studies of
galaxy formation and evolution.
The majority of LG galaxies are dwarfs, and dwarf galaxies are also the
most numerous type in the Universe.  Thus, dwarf galaxies have the potential to
provide important information on many key astrophysical processes.

In current structure formation models, dwarf galaxies are the first to
collapse, around 1 Gyr after the Big Bang, and thus their oldest
populations probe the conditions of the very early Universe. In
particular, the details of their early SFHs can provide information on the effects
on galaxy formation of the reionization which occurred at high
redshift. They also can shed light on the role of other mechanisms,
like Supernovae feedback, in removing significant amounts of gas from
small galaxy halos.

The MW satellites, the only galaxies which SFHs are known in detail,
may however not be good dwarf galaxy representatives for the above
purposes.  Because of the strong influence of our Galaxy, their
evolution is complicated by environmentally related mechanisms like
tidal stirring (Mayer et al. 2006) or ram pressure stripping (Blitz \&
Robishaw 2000), which are expected to remove gas and strongly affect
their SFHs.  In addition, the radiation field of the early MW is
expected to dominate over the cosmological UV background (Mayer et
al. 2005)
\begin{figure}[!ht]
\centerline{
\scalebox{0.65}{\includegraphics{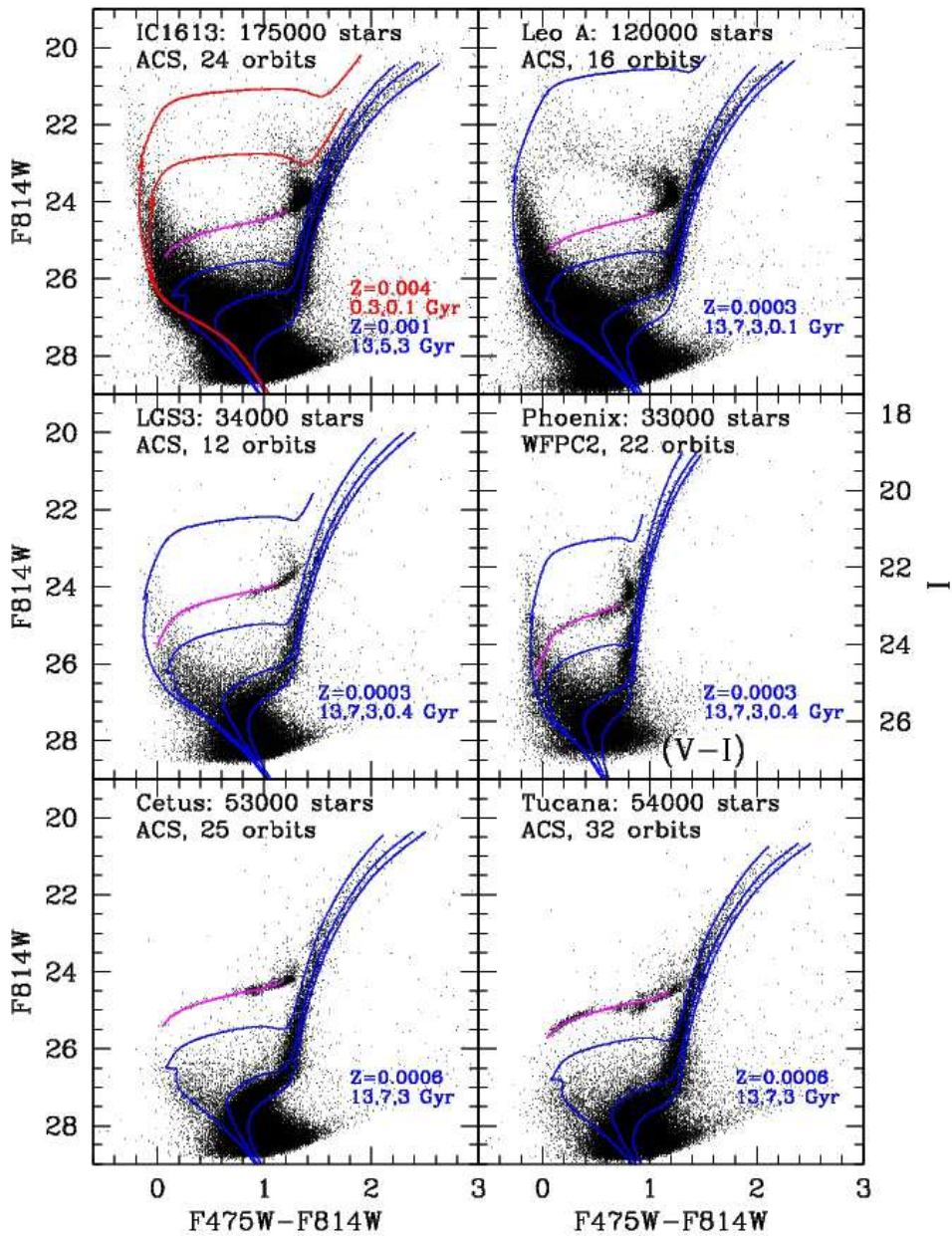}}}%
\caption{CMDs obtained for the six galaxies in our sample. 
In order to give a first indication of the range of ages and
metallicities present in each galaxy, isochrones from Pietrinferni et
al. (2004; overshooting set), with the ages and metallicities
indicated in the labels, have been superimposed. The locus of the
zero-age horizontal-branch (ZAHB) for Z=0.0006 or Z=0.0003 (depending
on the metallicity of the main population) is also shown.  Distance
moduli of $(m-M)_0=$ 24.5, 24.5, 24.1, 23.2, 24.5 and 24.8 and
reddenings of E(B-V)=0.04, 0.02, 0.05, 0.02, 0.03 and 0.03 for IC1613,
Leo A, LGS3, Phoenix, Cetus and Tucana, respectively, have been
adopted.  Note the different band combination in the case of
Phoenix. Determinations of the SFH of each system are under way
through comparison with synthetic CMDs.
\label{seis}}
\end{figure}

For these reasons, the smallest {\bf isolated} dwarf galaxies are
potentially interesting cosmological probes. Taking advantage of the great
sensitivity and spatial resolution of ACS on board the HST, we have
undertaken a major project aimed at investigating the SFH of a
selected sample of isolated LG dwarf galaxies. Our main goal is to
obtain CMDs reaching the oldest main-sequence turnoffs (M$_I \simeq
+3$ in low metallicity systems) with good photometric accuracy, for
six isolated dwarf galaxies (two dIrr galaxies: Leo~A and IC1613; the
two isolated dSph galaxies discovered so far in the Local Group: Cetus
and Tucana; and two transition type dIrr/dSph: LGS3 and Phoenix). The
data for this project have been obtained through two HST-ACS programs
(GO 10505, P.I. Gallart; GO 10590, P.I. Cole) with a total of 113
awarded orbits, and will be supplemented by data from a previous
HST-WFPC2 program for Phoenix, (GO 8706, P.I. Aparicio). The main aims
of this project are the following:

$\bullet$ To obtain the SFH of each galaxy through comparison of the
observed CMD with synthetic CMDs (see below).

$\bullet$ To study the radial gradients in the SFH for the galaxies in
the sample, in order to gain insight on the structural evolution
within dwarf galaxies.

$\bullet$ To characterize the populations of variable stars in each
galaxy, and to understand their relationships with the SFHs and stellar
population gradients. This will provide information on the use of
variable stars as stellar population tracers.
 
By comparing the results on isolated dwarfs with those obtained for
the MW and M31 satellites, we also expect to shed new light on the
effects of environmental mechanisms, which are expected to dominate
for the satellite sample. A detailed comparison between Cetus and
Tucana, and the dSph satellites will be particularly useful to clarify
to what extent these two groups of objects may be related to
each other.

Here, we will present some preliminary results regarding the obtained
CMDs and the study of the SFH gradients. Other papers in this
conference deal with complementary aspects, namely: the data reduction
strategy (Monelli), the population of variable stars in Tucana and
LGS3 (Bernard), and the first quantitative derivations of the SFH in
two of the galaxies: Leo A (Cole) and Phoenix (Hidalgo, Aparicio \&
Mart\'\i nez-Delgado).

\vspace{-0.38truecm}
\section{The SFH and stellar populations gradients from deep CMDs.}
\label{cmd}

In this project, we have emphasized the need to obtain CMDs reaching
the oldest main-sequence turnoff with good photometric accuracy, in
order to obtain reliable and detailed SFHs.  In this case, the
information can be obtained directly from the main sequence, which is
the best understood phase of stellar evolution from the theoretical
point of view and is also the one in which the location of stars in
the CMD shows the highest sensitivity to age differences.  The range
of ages and metallicities present can be determined simply through
comparison with theoretical isochrones. However, to
{\bf quantitatively} determine the SFH, it is necessary to compare the
observed density distribution of stars with that predicted by stellar
evolution models (see Gallart, Zoccali, \& Aparicio 2005).

Figure \ref{seis} shows the CMDs of all of the galaxies in our
sample. Note the variety of SFHs, as evidenced by the comparison with
selected isochrones. This is the first time that data of this high
quality have been obtained for dwarf galaxies beyond the MW satellite
system. The determination of the SFHs, through comparison of observed
and synthetic CMDs, is underway using different approaches by
different team members (see e.g., Aparicio
\& Gallart 2004; Bertelli \& Nasi 2001; Gallart et al. 1999; Skillman 
et al. 2003; and references therein, for a discussion of the different
approaches.  See also
http://iac-star.iac.es/iac-star/). This will allow us to evaluate
systematic differences in the results, and to better understand the
errors affecting the SFH determinations. See Cole (this conference, and
Cole et al. 2007) and Hidalgo et al. (this conference, and Hidalgo et
al. 2007) for the first results on Leo A and Phoenix respectively.

\begin{figure}
\centerline{
\scalebox{0.61}{\includegraphics{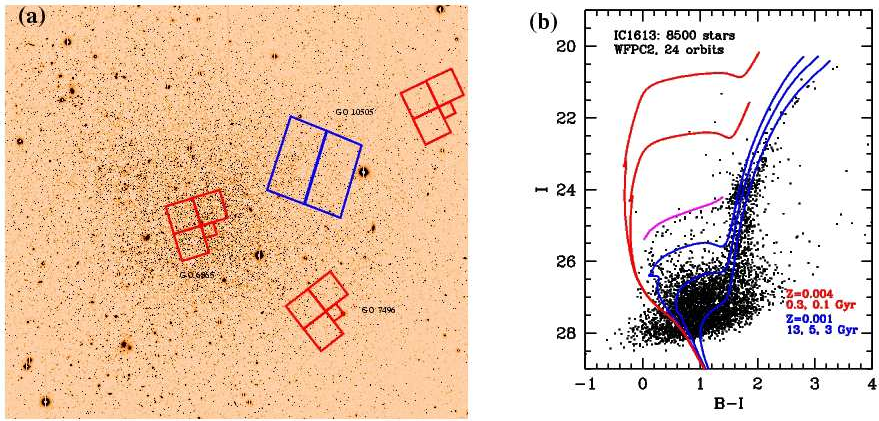}}}%
\caption{a) IC1613 field from a Subaru archival image. The location of our 
ACS and WFPC2 fields, and that of other WFPC2 pointings of varying depth is
shown. b) CMD of our WFPC2 parallel field, with the same isochrones as
in Figure~\ref{seis} superimposed.
\label{ic1613grad}}      
\end{figure}

We will also pay special attention to the study of the stellar
population gradients. In most cases, the ACS covers a significant
fraction of the galaxy, and therefore we will be able to determine a
detailed SFH as a function of galactocentric distance. This will
provide key information on the structural evolution of each galaxy
over a Hubble time, shedding new light on possible formation
mechanisms.  

In the case of IC1613, the ACS covers only a small fraction of the
galaxy, and we will rely on our WFPC2 parallel field at $\simeq$
12$^{\prime}$ from the center ({\it plus} several archival WFPC2
pointings and ground based data; see Figure~\ref{ic1613grad}a) to
obtain information on stellar population
variations. Figure~\ref{ic1613grad}b shows the CMD of IC1613 obtained
from our WFPC2 exposure. The same isochrones as in Figure~\ref{seis}
have been superimposed. Note the very little presence, if any, of
populations younger than $\simeq$ 2--3 Gyr. However, a substantial
amount of intermediate-age population (as young as 3 Gyr old) exists
even in this very peripheral position.

\begin{acknowledgments}
I thank my parents, Juan and Rosa, for coming with me and my little
son Daniel to yet another conference, thus allowing me to attend while
sharing a special time together.  Funding for this project is provided
by the IAC (grant P3/94), the Spanish Ministry of Education and
Science (grant AYA2004-06343, which includes EU FEDER funds), and NASA
(grants GO-10505 and GO-10590 from the STScI).
\end{acknowledgments}

\end{document}